# Brexit: Britain has spoken… or has it?

## Nicholas Donaldson, Nora Donaldson & Grace Yang

*Nicholas Donaldson*, *Nora Donaldson* and *Grace Yang* explore issues in the design and analysis of the Brexit referendum

The "Brexit" referendum on the United Kingdom's membership of the European Union (EU) took place on 23 June 2016. The result was essentially 52-48 in favour of Leave, as the observed proportion of Leavers was 51.9%. On this basis, statements like "the majority of the UK chose to leave the EU" or "the British people have voted to leave the European Union" or "the will of the British people is…" have pervaded political discourse and newspaper articles. However, all those statements are untrue or – at best – unproven.

If 100% of voters had turned out to vote, then there would have been complete information about the opinion of the UK voting population in the Brexit referendum, but this is not the case: the turnout rate was only about 72%, which means that only 37% of the eligible British electorate chose Leave. With turnout at 72%, Leave would needed to have won 69% of the vote to claim the majority of the British public was on its side with certainty. But since any statement about the whole population made on the basis of a subset carries with it an element of uncertainty, one should be compelled to perform the best possible analysis to deduce the views of the whole population from the subset. The uncertainty element should be part of the analysis.

In a recent paper, *What does the data of the Brexit referendum really say?*[1], the authors address two fundamental issues. Here, we summarize the discussion and add some context.

**An implied majority?**

The first issue is whether 51.9% is sufficiently larger than 50% to substantiate the claim that there was a majority for one of the campaigns, particularly the Leave campaign. If that were the case, we would say that the difference is of *practical* or *material* significance. *Practical/material significance*, in the case of the Brexit referendum, means that the difference is substantial and very likely represents a majority of the population. It is important here to mention that *practical/material significance* and *statistical significance* are not the same thing and one can have one without the other. Statistical significance refers to the strength of evidence against a hypothesized value carried by a sample. If the sample is large enough, statistically significant differences can be found even for differences that are of no practical/material significance and, conversely, if the sample is small enough, differences that are of practical and material significance can fail to reach statistical significance. The decision of where the threshold for material/practical significance should be depends upon the particular context and field, and should be made on the basis of appropriate specialized knowledge and understanding.

Common sense indicates that, a 60/40 split (i.e. a 10% difference over 50%), would provide a plausible foundation for a claim of a majority, and that a 50/50 split would have been a perfectly even split indicating a tie. And perhaps common sense is all that is required to realize that a 52/48 split represents a difference of no material or practical significance; that the observed difference of 2% (over 50%) is not sufficiently large as to imply an unquestionable majority and rather, conclusively, indicates a tie. This is precisely what was suggested in the public domain by two individuals, both supporters of the Leave campaign, in advance of the referendum, at the time when the polls were predicting such a 2% advantage for Remain. One was the political leader of the UK Independence Party (UKIP) and the other was a member of the public in an e-petition submitted to Parliament on May 25, 2016. The petition was later endorsed by more than 4 million people and stated: "We the undersigned call upon HM Government to implement a rule that if the remain or leave vote is less than 60% based a turnout less than 75% there should be another referendum." [2]

The petition was brought up in the House of Commons and rejected, and at the same time the Government confirmed that the terms of reference for Brexit had not stated the two crucial parameters of a referendum: the threshold for the result and the minimum turnout required to declare victory for either campaign. The Government wrote: "The European Union Referendum Act received Royal Assent in December 2015, receiving overwhelming support from Parliament. The Act did not set a threshold for the result or for minimum turnout." Omitting such essential terms of reference should not have happened in the first place. A referendum is a vote used to elucidate the opinion of a population in regard to a particular issue, but (to take an extreme example) it does not presuppose that a difference of 1 individual in the split should qualify as a majority in a population of almost 50 million. Or, conversely, a split of 100% vs 0% with a turnout of only 10 people should not qualify as a majority in such a huge population. The question that the Government should have considered in December 2015, at the planning stage, is: where should the threshold for the result and the minimum turnout be fixed to claim a majority for either side? In the absence of these terms of reference, after the Brexit referendum, the question is: does such a small difference represent material or practical significance?

In assessing whether a 2% difference is of *practical or material* significance, the following procedure is used. The difference is divided by the residual standard deviation in order to provide a *standardized* difference, which is also called the "effect size". The greater the standard deviation for a given difference, the smaller the associated effect size. This standardization allows the magnitude of the difference to be assessed, regardless of the units in which the responses were measured so that its true magnitude, in relation to the variation across the population, can be easily assessed. For normally distributed data, the effect size can be expressed in Cohen's *d* scale, in which the size of the difference can be judged, from negligible/very-small to large/very-large. In this scale, widely used in behavioral, social and biomedical sciences, an effect size less than 0.10 signals a very small effect and less than 0.05 enters the realm of negligible effects. Since the responses in the referendum are binary (Leave or Remain) and therefore not normally distributed, the proportion (of Leavers) is transformed into the logarithm of the odds (of Leave in relation to remain), which is close to normally distributed, and can be easily translated into Cohen's *d* effect size.

The split 48.1/51.9 in the Brexit referendum corresponds to an odds of choosing Leave (in relation to Remain) of 1.079. This means that the odds of a voter choosing to leave were about 8% greater than choosing to remain. The log-odds is 0.076 and the effect size expressed in the standardized Cohen's *d* metric is therefore $d=0.042$. This falls in the lowest quarter of the small range of the scale, indicating that the difference represented by 51.9% is of no material or practical significance. On this basis, the claim that the majority of the respondents of the UK voted in favour of leaving the EU is not substantiated by the 51.9% for Leave. Such a split indicates a tie between the two campaigns. In Donaldson & Yang's paper, the authors show that 60/40 is the minimum split required to justify a claim of a majority for either side, a split that is of any material or practical significance. Even a split of 55/45 would be considered a small difference that could be questioned if used to declare a majority for either side. It is relevant to note that in this part, because the Brexit voting population is so large (over 33 million), we consider only the issue of practical or material significance; not statistical significance.

**Differences between regions and countries**

The second fundamental issue addressed in Donaldson & Yang's paper has to do with the actual calculation of the proportion favouring Leave and Remain in the UK population. If significant variability between regions is detected, the mathematics used for the calculation of the proportion of Leave voters (or Remain voters) needs to be slightly more sophisticated than a simple arithmetic average of the aggregated population, to avoid bias in the results.[3]

In terms of the voting choice in the Brexit referendum, the population of the 326 *local authorities* in England exhibited a high degree of *granularity* (the extent of fine and exhaustive level of detail used by a model to treat a population). If the population is composed of many groups of indistinguishable units (or *grains*) in terms of a given outcome, the model would gain efficiency if the fine detail is averaged out over groups that are indistinguishable within and distinguishable between. A model stratifying England into the 533 electoral districts (constituencies) produced the same degree of granularity in relation to the exhaustive population described at the local authority level and, for this reason, Donaldson & Yang stratified England into 9 regions (North West, North East, East, Yorkshire, West Midlands, East Midlands, London, South West and South East), as established in 1994 to promote economic development. These regions showed voting behaviors that were essentially indistinguishable within regions and distinguishable between regions. Thus the whole UK population was stratified into these 9 regions for England and the three other countries.

While the UK is a sovereign state, it is not a homogeneous population: the four countries that comprise it have active cultural divisions. Three of the countries (Scotland, Northern Ireland and Wales), through devolution, have degrees of individual political authority. A research study by ETHNOS in 2005 saw that "Scottish and Welsh participants identified themselves much more readily as Scottish or Welsh rather than as British", an identity of which their English fellows used indistinguishably from Englishness.[4] As well as national identity, within each of these countries different regions exhibit variation in age, income, and education. All of this suggests economic, political, and cultural diversities that demonstrate the heterogeneity of the UK in relation to the EU issue. Acknowledging heterogeneity should have constituted a preliminary step in planning the referendum and the calculations should take into account, as well as the size of the regions, the within-region variability and the between-region heterogeneity that was present in the UK population.

The heterogeneity assessment is based on Cochran's Q statistic, whose distribution (for our data) can be approximated by a Chi-square distribution with $k-1$ degrees of freedom (d.f.), where $k$ is the number of regions (or countries) that intervene in the calculation; the mean of the Chi-square distribution is the number of degrees of freedom, $k-1$ in this case. In Donaldson & Yang's paper, the authors first demonstrated that there is a highly significant heterogeneity between the 13 geo-political regions of the UK – Wales, Scotland, Northern Ireland, Gibraltar and the 9 regions of England – (Cochran's $Q(12\ df)$=639,057; P<0.0001) and also between the 9 regions of England (Cochran's $Q(8\ df)$=383,792; P<0.0001). The actual proportion of Leavers is then calculated using meta-analytic algorithms that control for heterogeneity. Three different models are used for comparison. The first is a *fixed effects* (FE) model, which ignores between-region heterogeneity but gives more weight to those regions with larger sample size and with smaller variability (i.e. more precision); it weights the parameter in each region by the inverse of its variance.[3] If there is no significant between-region heterogeneity, as a minimum, a FE algorithm should be used. Subsequently, two models are presented that control for between-region heterogeneity. One is the widely-used random effects (RE) model, which aims to improve on the FE model by controlling for the between-region heterogeneity. To do this, it uses weights involving the inverse of the total variance (within-region variance plus between-region variance).[5] The other is a fixed-effects model that controls for region heterogeneity (IV-Het)[6] providing an improvement over the RE model in that, when the between-region variability is substantial, the weights will not migrate towards uniformity.

**The results in England.** Out of the nine regions of England, we identified three subgroups that, while still heterogeneous (P<0.001), were much less so than the 9-region partition: Cochran's Q statistic for each group was of the order of 1% of that for the 9-region partition. The first group is formed by five regions: the North East, East, Yorkshire, West Midlands and East Midlands. With a Cochran's $Q(4\ df)$=6,062, their effect sizes were between 0.14 and 0.20, a *small size advantage* for Leave. The second group was formed by three regions: North West, South West and South East. With a Cochran's $Q(2\ df)$=2,743, their effect sizes were between 0.06 and 0.08, *a very small size* advantage for Leave. The third group was London; its effect size was $d$=-0.20, a *medium size* advantage for Remain. With this partition, the FE model controlling for heterogeneity (IV-Het), giving weights of 44.6%, 41.7% and 13.7% to the first, second and third regions respectively, yield an overall effect of $d$=0.13, a small size advantage for Leave. The RE model yields an overall effect of $d$=0.01, a negligible advantage for Leave. In both cases, the 95% confidence intervals include zero: the advantage for Leave in England is not found to be statistically significant.

**The results in the UK.** When the UK is divided into its four countries (England, Northern Ireland, Scotland and Wales), with England partitioned into the three subgroups, the heterogeneity is considerably reduced but still highly significant (Cochran's $Q(5\ df)$=228,257; P<0.001). The random effects (RE) model yields an overall effect of $d$=-0.10, which represents a small advantage for Remain. The fixed effects model controlling for heterogeneity (IV-Het) gives weights of 37.7% to Yorkshire-Midlands-E-NE, 35.3% to NW-SE-SW, 11.5% to London, 7.4% to Scotland, 5.1% to Wales and 2.9% to Northern Ireland. It yields an overall effect of $d$=0.08, which in Cohen's scale represents a small advantage for Leave. In both cases the 95% confidence intervals include zero: under both models the effect is not found to be statistically significant.

**Summary and discussion.**

Two fundamental issues with the Brexit referendum lie in the data analysis adopted and in the lack of fixing a threshold for the result and the minimum turnout. Effect size and practical/material significance seem to be concepts that have not made their way into the toolkit of most political scientists or politicians. In contrast, there is consensus among behavioural, social, and biomedical scientists that effect sizes should always be reported; that they are relevant even in cases when statistical hypothesis tests have to be abandoned altogether, and that the concept of *material/practical significance* cannot be disregarded. If only these two concepts had been employed at the design stage, the terms of reference would have reflected the divisions of the United Kingdom and the result, either way, being for Leave or Remain, would have been accepted more readily. Members of government are elected to perform jobs which are often complex and require the collection and evaluation of evidence. To do this correctly, they should seek advice from the relevant experts. On this basis, a relevant question is: "Did the Government take advice from political scientists and statisticians in December 2015?" And, most crucially: *"Did the Government take advice from political scientists and statisticians in 2016 when they debated the issues of design and analysis raised in the petition endorsed by over 4 million people?"*

All algorithms used in Donaldson & Yang's paper consistently suggest that the result of the Brexit referendum indicates no significant advantage for either side. The better the adjustment for the highly significant heterogeneity, the further the aggregation moves away from the Leave side, while still showing that there is a tie between the two campaigns. Donaldson & Yang find that the more the heterogeneity present in the Brexit referendum is adjusted for, the further the results move away from Leave, albeit still within the range of small standardized differences. In all cases, the differences are consistently found to be of, essentially, no practical (or material) significance. Therefore, the data of the Brexit referendum does not support a claim of majority for either side; it signals a tie between the two campaigns. The rejection of the petition[2] for a second referendum by the Government, and the reason given to support the rejection (i.e. that the European Union Referendum Act did not set a threshold for the result or for minimum turnout), amount to a denial of the fact that the result of the Brexit referendum was, essentially, a tie.

Nicholas Donaldson is a freelance journalist and research assistant with strong interest in social science, cultural marketing, ethnography and communications. He previously worked for Latest TV Brighton as a video journalist.

Nora Donaldson is Honorary Professor of Biostatistics at King's College Hospital NHS Foundation Trust (UK) and Affiliate and previously Professor of Biostatistics at King's College London (University of London).

Grace Yang is Emeritus Professor in Statistics at the University of Maryland at College Park, US.